\documentclass[12pt]{article}
\usepackage{amsmath,amsfonts,xypic, epsfig}
\usepackage{amssymb}
\usepackage{amscd}
\usepackage[all]{xy}
\advance \textheight by \topskip
\makeatletter

\@addtoreset{equation}{section}
 \def\theequation{\thesection.\arabic{equation}}
\makeatother

\newtheorem{theorem}{Theorem}[section]

\newtheorem{lemma}[theorem]{Lemma}

\newcommand{\qed}{{\hfill $\Box$}}

\def\NN{{\mathbb N}}

\def\RR{{\mathbb R}}
\def\NN{{\mathbb N}}

\def\ln{{\mathrm{ln}}}
\newcommand{\beqa}{\begin{eqnarray}}
\newcommand{\eeqa}{\end{eqnarray}}
\newcommand{\noi}{\noindent}

\def\>{\rangle}
\def\<{\langle}

\begin{document}

\title{
{\bf A translation-invariant renormalizable non-commutative scalar model
}  
\author{ 
{\sf  R. Gurau${}^{a}$, J. Magnen${}^{b}$, V. Rivasseau${}^{a}$ and A. Tanasa${}^{c,d}$}
\\
{\small ${}^{a}${\it Laboratoire de Physique Th\'eorique, CNRS UMR 8627,}} \\
{\small {\it b\^at. 210, 
Universit\'e Paris XI, 91405 Orsay Cedex, France}}  \\ 
{\small ${}^{b}${\it Centre de Physique Th\'eorique, CNRS UMR 7644,}} \\
{\small {\it Ecole Polytechnique, 91128 Palaiseau, France}}  \\ 
{\small ${}^{c}${\it Institutul de Fizica si Inginerie Nucleara Horia Hulubei,}} \\
{\small {\it P. O. Box MG-6, 077125 Bucuresti-Magurele, Romania}}  \\ 
{\small ${}^{d}${\it Max-Planck-Institut fur Mathematik,}} \\
{\small {\it Vivatsgasse 7, 53111 Bonn, Germany}}  \\ 
}}
\maketitle
\vskip-1.5cm

\vspace{2truecm}

\begin{abstract}
\noindent
In this paper we propose a  translation-invariant scalar model on the Moyal space. We prove that this model does not suffer from the UV/IR mixing and we establish its renormalizability to all orders in perturbation theory.
\end{abstract}


\section{Introduction and motivation}
\renewcommand{\theequation}{\thesection.\arabic{equation}}   
\setcounter{equation}{0}

Space-time coordinates should no longer commute at the Planck scale where gravity should be quantized.
This observation is a strong physical motivation
for non-commutative geometry, a general
mathematical framework  developed by A. Connes and others \cite{book-connes}.
Non-commutative field theory is the reformulation of ordinary quantum
field theory on such a non-commutative background. It may represent
a bridge between  the current standard model of quantum fields
on ordinary commutative $\RR^4$ and a future formalism including quantum gravity
which hopefully  should be background independent.

Initially there was hope that  non commutative field theory would behave better in the ultraviolet regime \cite{Snyder}.
Later motivation came from string theory, because 
field theory on simple non-commutative spaces (such as 
flat space with Moyal-Weyl product) appear
as special effective regimes of the string \cite{string1,string2}. 
Finally an other very important motivation comes from the study of
ordinary physics in strong external field (such as the quantum Hall effect) \cite{hall1,hall2,hall3}. Such situations
which have not been solved analytically with the ordinary commutative techniques may
probably be studied more fruitfully with non-commutative techniques.

Renormalization is the soul of ordinary field theory and one would certainly want to
extend it to the non-commutative setting. But
the simplest non-commutative model, namely $\phi^{\star 4}_4$, whose action is given
by (\ref{act-normala}) below,
was found to be not renormalizable because of a surprising phenomenon 
called \emph{UV/IR mixing}  \cite{melange}. This mixing also occurs in non-commutative Yang-Mills
theories. Roughly speaking the non-commutative theory still has infinitely many ultraviolet divergent graphs 
but fewer than the ordinary one. However some ultraviolet
convergent two point graphs, such as the "non-planar tadpole"  
generate infrared divergences which are not of the renormalizable 
type \footnote{This UV/IR mixing although quite generic may 
be avoided in some classes of "orientable models". 
Remark also that in Minkowski space if one maintains a rigorous notion of causality, there
are strong indications that ultraviolet/infrared mixing does not occur
\cite{Bahns}. However the Minkowski theory has
complications of its own which make it harder to study.}.

The first path out of this riddle came when
H. Grosse and R. Wulkenhaar introduced a modified $\phi^{\star 4}_4$ model which is renormalizable \cite{GW1,GW2}.
They added  to  the usual propagator a marginal harmonic potential, 
which a posteriori appears required by 
Langmann-Szabo duality $\tilde x_{\mu} = 2 \theta_{\mu \nu} x^{\nu}  \leftrightarrow p_{\mu}$
\cite{LaSz}.

The initial papers were improved and confirmed over the years through several independent
methods \cite{RVW,GMRV}. The main property of the Grosse-Wulkenhaar model is that its 
$\beta$-function vanishes at all orders at the self-duality point $\Omega=1$ \cite{beta1,beta23,beta}. 
The exciting conclusion is that this model is asymptotically safe,
hence free of any Landau ghost, and should be a fully consistent theory
at the constructive level. This is because wave function renormalization exactly compensates
the renormalization of the four-point function, so that the flow
between the bare and the renormalized coupling is bounded.

Essentially most of the standard tools of field theory
such as parametric 
\cite{param1,param2}  
and Mellin representations,
\cite{mellin}
dimensional regularization and renormalization 
\cite{dimreg} 
and the Connes-Kreimer Hopf algebra formulation of renormalization 
\cite{hopf}
have now been generalized to renormalizable non-commutative quantum field theory. Other renormalizable models have been also developed 
such as the orientable Gross-Neveu model \cite{fab}.

For a general recent review on non-commutative physics
including these new developments on non-commutative field theory, see \cite{QS}.

However there are two shortcomings of the Grosse-Wulkenhaar (GW) model.
Firstly it breaks translation invariance so that its relevance to physics
beyond the standard model would be indirect at best;
one should either use more complicated "covariant" models with harmonic potentials which are invariant under
"magnetic translations", such as the Langmann-Szabo-Zarembo model \cite{LSZ}
or one should understand how many short distance localized GW models 
may glue into a translation-invariant effective model.
Secondly it is not easy to generalize the GW method to gauge theories,
which do present ultraviolet/infrared mixing. Trying to maintain both gauge invariance
and Langmann-Szabo duality one is lead to theories with non trivial vacua\cite{gauge1,gauge2, ultimul,goldstone}, 
in which perturbation theory is difficult and renormalizability to all orders is therefore unclear up to now.

Motivated by these considerations we explore in this paper an other solution to the 
ultraviolet infrared mixing for the $\phi^{\star 4}_4$ theory. It relies on the very natural idea to incorporate
into the propagator the infrared mixing effects. This is possible because the sign of the mixing graphs is the right one.
One can therefore modify the propagator to include from the start a  $1/p^2 $ term besides
the ordinary $p^2$ term, and to define new renormalization scales accordingly. 
Adding the interaction and expanding into the coupling constant
we prove in this paper that the model modified in this way is indeed renormalizable at 
all orders of perturbation theory.
This is because the former infrared effects now just generate a flow (in fact a finite flow) for
the corresponding  $1/p^2 $  term in the propagator.
The "ordinary"  $\phi^{\star 4}_4$ is formally recovered in the case where the
bare coefficient of the $1/p^2 $ term is zero.

The advantages of this "$1/p^2 - \phi^{\star 4}_4$" model are complementary to those of the GW model. 
The main advantage is that the model does not break translation invariance. The main inconvenient
is that there is no analog of the Langmann-Szabo symmetry so that one should not
expect this $\phi^{\star 4}_4$ model to make sense non perturbatively. However the real interest of this
work is perhaps to offer an alternative road to the solution of ultraviolet/infrared mixing
in the case of gauge theories. It may
lead to gauge and translation invariant models with trivial vacua.
Remark that since ordinary 
non abelian gauge theories are asymptotically free,
there is no real need for the non commutative version to behave better than the commutative case.
This removes some of the motivation to implement Langmann-Szabo duality in that case.
Therefore we hope the model studied here may be a step towards a better global proposal
for a non-commutative generalization of the standard model. This proposal may perhaps have to combine
different solutions of the ultraviolet/infrared mixing in the Higgs and gauge sectors
of the theory.

The paper is organized as follows. Section \ref{sectionmodel}  recalls useful facts about Feynman graphs.
and defines our model. The main result of the paper, theorem \ref{mainres} below is stated
at the end of that section.
The proof is through the usual renormalization group multiscale analysis.
The definition of the renormalization group slices and the power counting
is given in section \ref{slicpower} and the proof of the theorem is completed in section \ref{renorm}
using the momentum representation.
Finally some low order renormalized amplitudes for this theory are 
computed in Appendix \ref {exemple}.

\section{Model and Main Result} \label{sectionmodel}
\renewcommand{\theequation}{\thesection.\arabic{equation}}   
\setcounter{equation}{0}

\subsection{ The ``naive''  $\phi^{\star 4}$ model}
It is obtained by replacing the ordinary commutative action 
 by the Moyal-Weyl $\star$-product
\beqa
\label{act-normala}
S[\phi]=\int d^4 x (\frac 12 \partial_\mu \phi \star \partial^\mu \phi +\frac
12 \mu^2 \phi\star \phi  + \frac{\lambda}{4!} \phi \star \phi \star \phi \star \phi),
\eeqa
with Euclidean metric. The commutator of two coordinates is
\beqa
[x^\mu, x^\nu]_\star=\imath \Theta^{\mu \nu},
\eeqa
where
\beqa
\label{theta}
\Theta=
\begin{pmatrix}
   0 &\theta & 0 & 0\\   
   -\theta & 0  & 0 & 0\\
   0&0 & 0 & \theta\\  
   0& 0 & -\theta & 0  
  \end{pmatrix}.  
\eeqa
In momentum space the action \eqref{act-normala} becomes
\beqa
\label{act-normala-p}
S[\phi]=\int d^4 p (\frac 12 p_\mu \phi  p^\mu \phi +\frac
12 \mu^2 \phi  \phi  + \frac{\lambda}{4!} \phi \star \phi \star \phi \star \phi).
\eeqa
The propagator is the same as in the commutative case
\beqa
\label{propa-normal}
\frac{1}{p^2+\mu^2}.
\eeqa

\subsection{Feynman graphs: planarity and non-planarity, rosettes}
\renewcommand{\theequation}{\thesection.\arabic{equation}}   
\setcounter{equation}{0}
\label{Feynman}

In this subsection we give some useful conventions and  definitions.
Consider a $\phi^{\star 4}$ graph with $n$ vertices, $L$ internal lines and $F$
faces. 
One has
\beqa
\label{genus}
2-2g=n-L+F,
\eeqa
\noi
where $g\in\NN$ is the {\it genus} of the graph.
If $g=0$ the graph is {\it planar}, if $g>0$ it is {\it non-planar}. 
Furthermore, we call a planar graph {\it regular} if it has a single face broken by external lines. 
We call $B$ the number of such faces broken by external lines.

The $\phi^4$ graphs also obey the relation
\beqa
\label{alta}
L=\frac12 (4n-N),
\eeqa
where $N$ is the number of external legs of the graph.

\begin{figure}
\centerline{\epsfig{figure=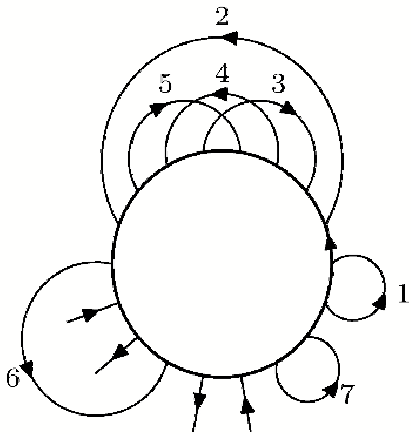,width=6cm} }
\caption{An example of a rosette}\label{roz}
\end{figure}

In \cite{filk}, T. Filk defined "contractions moves" on a Feynman 
graph. The first such move consists in reducing a tree line and gluing
together the two vertices at its ends into a bigger one.
Repeating this operation for the $n-1$ lines of a tree, one obtains a single final
vertex with all the loop lines hooked to it - a {\it rosette} (see Fig. \ref{roz}).

Note that the number of faces and the genus of the graph do not change 
under this operation. Furthermore, the external legs will break the same faces on the rosette. When one deals with a planar graph, there will be no crossing between the loop lines on the rosette. The example of Fig. \ref{roz} corresponds thus to a non-planar graph (one has crossings between e.g. the loop lines $3$ and $5$). 
Following \cite{filk} the rosette amplitude  is
\beqa
\label{contributia}
\tilde V (\mbox{external momenta}) \, e^{\frac \imath2 \sum_{ij} I_{ij}
  \Theta_{\mu\nu}k_i^\mu k_j^\nu}
\eeqa
where the intersection matrix $I_{ij}$  is given by
\beqa
\label{intersectia}
I_{ij}=
\begin{cases}
1, \mbox{ if line $j$ crosses line $i$ from right,}\\
-1  \mbox{ if line $j$ crosses line $i$ from left,}\\
1  \mbox{ if lines $i$ and  $j$ do not cross,}
\end{cases}
\eeqa
where $i$ and $j$ correspond to an (arbitrary) numeration of the lines,
independent of them being external or internal lines of the Feynman graph. An
orientation is given by the sign convention chosen for the momenta in the
conservation conditions. $\Theta$ is the non-commutativity matrix (see equation
\eqref{theta}). Furthermore the overall phase factor corresponding to the
external momenta is
\beqa
\label{overall}
\tilde V(k_1, \ldots, k_N) = \delta (k_1+\ldots k_N) e^{\frac \imath2 \sum_{i<j}^N
  k_i^\mu k_j^\nu \Theta_{\mu \nu}} ,
\eeqa
which has exactly the form of a Moyal kernel.

\subsection{UV/IR Mixing}

\begin{figure}[ht]
\centerline{\psfig{figure=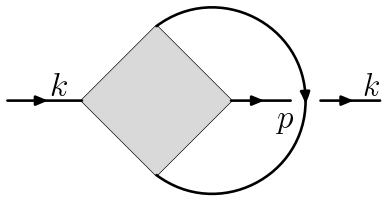,width=5cm}}
\caption{The non-planar tadpole}
\label{fig:tadpole}
\end{figure}

The non-locality of the $\star$-product leads to a new type of divergence, the
UV/IR mixing \cite{melange}. This can be  seen already in
the non-planar tadpole (see Fig. \ref{fig:tadpole}). Although
this graph has zero genus, since it has two faces broken by external
lines, it will lead to non-planarity when inserted into larger graphs.

The amplitude of this non-planar tadpole with internal momentum higher than the external momentum is up to a constant
\beqa
\label{tadpole}
T=\int_{0}^{k^{-2}} d\alpha \int d^4 p e^{\imath k\Theta p}e^{-\alpha(p^2+\mu^2)}.
\eeqa
Integrating the Gaussian (and setting $\theta=1$) holds
\beqa
T=\int_{0}^{k^{-2}} \frac{d\alpha}{\alpha^2} e^{- \frac{k^2}{\alpha}}e^{-\alpha \mu^2}.
\eeqa
If $k>1$ then 
\beqa
|T|<\int_{0}^{\infty} \frac{d\alpha}{\alpha^2}e^{-\frac{1}{\alpha}}=1\, .
\eeqa

Let $k<1$ we have
\beqa
T=\int_{0}^{k^{2}} \frac{d\alpha}{\alpha^2} e^{- \frac{k^2}{\alpha}}e^{-\alpha \mu^2}+
\int_{k^2}^{k^{-2}} \frac{d\alpha}{\alpha^2} e^{- \frac{k^2}{\alpha}}e^{-\alpha \mu^2},
\eeqa
Rescaling $\alpha=k^2\beta$ we have for the first integral 
\beqa
\frac{1}{k^2}\int_0^1 \frac{d\beta}{\beta^2}e^{-\beta^{-1}}e^{-\beta k^2 m^2}=\frac{F(k)}{k^2} \, ,
\eeqa
with $F$ an analytic function of $k$.

We separate again the second integral as
\beqa
\int_{k^2}^{\mu^{-2}} \frac{d\alpha}{\alpha^2} e^{- \frac{k^2}{\alpha}}e^{-\alpha \mu^2}+
\int_{\mu^{-2}}^{k^{-2}} \frac{d\alpha}{\alpha^2} e^{- \frac{k^2}{\alpha}}e^{-\alpha \mu^2}.
\eeqa
The second integral is bounded by a constant uniformly in $k$. In the first integral we 
Taylor-expand $e^{-\alpha \mu^2}$ to get
\beqa
\int_{k^2}^{\mu^{-2}} \frac{d\alpha}{\alpha^2} e^{- \frac{k^2}{\alpha}}-
\int_{k^2}^{\mu^{-2}} \frac{d\alpha}{\alpha^2} e^{- \frac{k^2}{\alpha}} \alpha \mu^2+
\int_{k^2}^{\mu^{-2}} \frac{d\alpha}{\alpha^2} e^{- \frac{k^2}{\alpha}} O(\alpha^2).
\eeqa
The first term integrates to $k^{-2}F'(k)$ with $F'$ analytic, the second  computes to $ \mu^2 \ln k^2+F''(k)$ with $F''$ analytic and the third is uniformly bounded. Thus the tadpole is
\beqa
T=\frac{c}{k^2}+c'\ln(k^2)+F(k)
\eeqa
with $c$ and $c'$ constants and $F$ an analytic function at $k=0$.

We note that for a non-massive model the second term vanishes. One can include 
the contribution of the non-planar tadpole in the complete  two-point function to obtain a dressed propagator.
This motivates a modification of the
kinetic part of the action \eqref{act-normala-p}
which leads to

\subsection{The $1/p^2$ $\phi^{\star 4}_4$ Model}
This model is defined by the following action
\beqa
\label{revolutie}
S[\phi]=\int d^4 p (\frac 12 p_{\mu} \phi  p^\mu \phi  +\frac
12 \mu^2  \phi  \phi   
+ \frac 12 a  \frac{1}{\theta^2 p^2} \phi  \phi  
+ \frac{\lambda }{4!} \phi  \star \phi  \star \phi  \star \phi ).
\eeqa
with  $a $ is  some dimensionless parameter.
The propagator is
\beqa
\label{propa-rev}
\frac{1}{p^2+\mu^2+\frac{a}{\theta^2 p^2}} \, ,
\eeqa
and we choose $ a \ge 0 \, $
so that this propagator is well-defined and positive.

Using \eqref{contributia} and \eqref{propa-rev} the amplitude of a $N$-point graph writes
\beqa
\label{total}
{\cal A}(G)&=&\delta (\sum_{i=1\dots N} k_i)e^{\frac \imath2 \sum_{i<j}^N
  k_i \Theta k_j}
 \int  \prod_{i=1}^L d^4 p_i \frac{1}{p_i^2+\frac{a}{\theta^2 p_i^2}+\mu^2}
\nonumber\\
&&\prod_{v\neq \bar v}\delta(q^v_1+q^v_2+q^v_3+q^v_3)
 e^{\frac \imath2 \sum_{ij} I_{ij}
  q^v_i\Theta q^v_j},
\eeqa
with $k$ the external momenta, $p$ the internal momenta, $q^v$ a generic notation for internal and external momenta
at vertex $v$, and $\bar v$ an external vertex of the graph chosen as root (to extract the global $\delta$ conservation
on external momenta).

The main result of this paper is
\begin{theorem}[Main Result] \label{mainres}
The model defined by action (\ref{revolutie}) is perturbatively renormalizable to all orders.
\end{theorem}
The proof is given in the next two sections.
We proceed now to the usual RG analysis by defining slices and establishing power counting.

\section{Slices and Power counting}  \label{slicpower}
\renewcommand{\theequation}{\thesection.\arabic{equation}}   
\setcounter{equation}{0}

In this section we establish the power counting of our model.
For that purpose we shall use the very powerful tool
of multiscale analysis. Power counting and renormalization
theory rely on some scale decomposition and renormalization group is {\it oriented}: it integrates 
"fluctuating scales" (which we call here "high" scales) and computes an effective action 
for background scales (here called "low" scales). There are several technical different ways 
to define the RG scales, but in perturbation theory the best way is certainly to
define the high scales as the locus where the denominator $D$ of the propagator is big
and the low scales as the locus where it is small, cutting the slices into a geometric progression.
This certainly works well for the very different RGs of ordinary statistical mechanics ($D= p^2$), of
condensed matter ($D= ip_0 + (\vec p)^2 -1$) and of the Grosse-Wulkenhaar model ($D= p^2 + \Omega^2 x^2$). 
We use the same idea here again with $D= p^2 + a/p^2$.

Power counting then 
evaluates contributions of connected subgraphs, also called "quasi local components" for which all internal scales
are "higher" than all external scales in the sense above. We shall not rederive this basic principle here and shall use 
directly the particular version and notations of \cite{carte}, in which these quasi-local components
are labeled as $G^j_r$.

Before going into the detailed analysis of this contributions we first note a very important feature of our model: the term $ap^{-2}$ changes the UV and IR regions. For the rest of this paper we set $\theta =1$. 
We employ the Schwinger trick and write:
\beqa
\frac{1}{p^2+a p^{-2}+\mu^2}=\int_{0}^{\infty}e^{-\alpha (p^2+a p^{-2}+\mu^2)} d \alpha \, ,
\eeqa

Let $M>1$. Slice the propagators as
\beqa
\label{slicebound}
C(p)&=&\sum_{i=0}^{\infty}C^i(p),\nonumber\\
 C^i(p)&=&\int_{M^{-2i}}^{M^{-2(i-1)}}
d\alpha e^{-\alpha (p^2+ a p^{-2} +\mu^2)} \le
Ke^{-c M^{-2i}(p^2+a p^{-2} +\mu^2)}, \,   i\ge 1, \nonumber
\\
C^0(p) &=&\int_{1}^{\infty}
d\alpha e^{-\alpha (p^2+ a p^{-2} +\mu^2)} \le
Ke^{-c p^2} \, ,
\eeqa
with $K$ and $c$ some constants which, for simplicity, will be omitted from now on. 
To the $i$-th slice corresponds either a momentum $p\approx M^{i}$ or a momentum $p\approx M^{-i}$. 
Conversely, a momentum $k\approx M^e$ for $e \in {\mathbb Z}$ has a scale $\alpha=M^{-2|e|}$.

We have the following lemma:

\begin{lemma}
The superficial degree of convergence $\omega (G)$ of a Feynman graph $G$ corresponding to the action
\eqref{revolutie} obeys
\beqa
\label{power}
\omega (G)\ge\begin{cases}
				N(G)-4, & \mathrm{ if }~g(G)=0\\
				N(G)+4 & \mathrm{ if }~g(G)>0\\
             \end{cases}
\eeqa
where $N(G)$ is the number of external legs of $G$, and $g(G)$ its genus.
\end{lemma}
\noi {\bf Proof}
The first line is easy. It is enough to take absolute values in (\ref{total}), and apply the momentum routing. We briefly recall this procedure. We fix a scale attribution for all propagators. As the sum over the scales is easy to perform 
(along the same lines as in \cite{carte}) 
we concentrate on the problem of summing the internal momenta at {\it fixed} scale attribution $\nu$.

At any scale $i$ the graph $G^i$ made of lines with scales higher or equal to $i$
splits into $\rho$ connected components $G^i_r$, $r=1,\dots, \rho$. We choose a spanning tree $T$ compatible with the scale attribution, that is each $T^i_r=T\cap G^i_r$ is a tree in the connected component $G^i_r$. We define the branch $b(l)$ associated to the tree line $l$ as the set of all vertices such that the unique path of lines connecting them to the root contains $l$.
We can then solve the delta functions for the tree momenta as
\beqa
p_l=-\sum_{l'\in b(l)}q_{l'}\,
\eeqa
where $l'\in b(l)$ denotes all loops or external momenta touching a vertex in the branch $b(l)$. After integrating
internal momenta  we get the bound
\beqa
{\cal A}^{\nu}\le \prod_{l} M^{-2i_l}\prod_{l\in {\cal L}}M^{4i_l} \, ,
\eeqa
where ${\cal L}$ denotes the set of loop lines. The first factor comes from the prefactors of the propagators while the second comes from the integration of the loop momenta. We can reorganize the above product according to the scale attribution as
\beqa
\label{powercounting}
{\cal A}^{\nu}\le \prod_{i,k}M^{-2L(G^i_r)}M^{4[L(G^i_r)-n(G^i_r)+1]}=
 \prod_{i,k}M^{-[N(G^i_r)-4]} \, ,
\eeqa
where we have used \eqref{alta}.

The second line of (\ref{power}) is obtained using an argument similar to the one used in \cite{GMRV}. 
In fact if the graph is non-planar there will be two internal loop momenta $p$ and $q$ such that, after integrating all tree momenta 
with the delta functions, the amplitude contains a factor
\beqa
I=\int d^4p\; d^4q \; e^{-\alpha_1 p^2- \alpha_1  a  p^{-2}
                       -\alpha_2 q^2- \alpha_2  a  q^{-2}+{\imath} p\wedge q} \ .
\eeqa
A naive bound would be to bound the integral by $M^{4i_1}M^{4i_2}$. Instead we use
\beqa
\frac{1}{(1+M^{2i_1}q^2)^m}\left(1+M^{2i_1}\sum_j\frac{d^2}{dp_j^2}\right)^m
e^{\imath p\wedge q}=e^{\imath p\wedge q} \, .
\eeqa
Integrating by parts we get
\beqa
\vert I \vert \le \int \frac{d^4pd^4q} {(1+M^{2i_1}q^2)^m}
e^{ -M^{-2i_2}q^2-M^{-2i_2} q^{-2}}
\left(1+M^{2i_1}\sum_j\frac{d^2}{dp_j^2}\right)^m
e^{-M^{-2i_1}p^2-M^{-2i_1}  p^{-2}   } \, .
\eeqa
The derivative acting on the exponential gives factors of order at most $O(1)$.  If we chose $m=3$ we have a bound
\beqa
I\le K\int \frac{d^4pd^4q} {(1+M^{2i_1}q^2)^3}e^{-M^{-2i_1}p^2}\le K'
\eeqa
We have thus gained both factors $M^{4i_1}$ and $M^{4i_2}$ with respect to the naive bound.
\qed

\section{Renormalization}\label{renorm}
\renewcommand{\theequation}{\thesection.\arabic{equation}}   
\setcounter{equation}{0}

We have established that all possible divergences come from planar $2$ or $4$ point graphs. Note that they may have more that one broken face\footnote{This stands in contrast with the Grosse-Wulkenhaar theory in which only graphs with a single broken face diverge.}.
We will prove that all divergences can be reabsorbed in a redefinition of the parameters in the action (\ref{revolutie}).

\subsection{Two-point  function}

The single-broken-face $2$-point graphs are ultraviolet divergent and as such give nontrivial mass and wave function renormalizations. By contrast the $2$-point graphs with two broken faces are ultraviolet convergent. Nevertheless we will prove 
that they give a finite renormalization of the $1/p^2$ term. 

\subsubsection{$2$-point function with a single broken face }

From the standard multiscale analysis we know that power counting has to be computed
only for connected components of the $G^j_r$ type. Consider the case of such a planar, one particle irreducible, $2$-point subgraph $S$ which
is  a component  $G^j_r$ for $j$ for a certain range of slices $e < j \le i$ between $e$, its highest external scale
and $i$, its lowest internal scale (and a particular value of $r$).
\beqa
{\cal A}(G^j_r)=\delta (k_1+k_2)
 \int  \prod_{l=1}^L d^4 p_l \int^{M^{-2(i_l-1)}}_{M^{-2i_l}}d\alpha_l~e^{-\alpha_l [p_l^2+ a p_l^{-2}+\mu^2]}
\prod_{v\neq \bar v}\delta(q^v_1+q^v_2+q^v_3+q^v_3) \, ,
\eeqa
where we consider that all eventual subrenormalizations have been performed. We perform the momentum routing for the subtree $T_r^j$. Let $k_1$ enter into the root vertex of $S$. We define
\beqa
T^1=\{l\in T~|~k_2\in b(l)\} \ , \quad T^2=T-T^1 \, .
\eeqa
The amplitude writes, dropping the index on $k_2$ and forgetting the overall $\delta$ function
\beqa
&& {\cal A}(G)= \int  \prod_{l=1}^{|{\cal L}|} d^4 p_l 
\prod_{l=1}^{L}  \int^{M^{-2(i_l-1)}}_{M^{-2i_l}} d\alpha_l
~\prod_{l\in {\cal L}} e^{-\alpha_l[p_l^2+a p_l^{-2}+\mu^2]}\nonumber\\
&&\prod_{l\in T^2}
e^{-\alpha_l\Big{[}(\sum_{l'\in b(l)}p_{l'})^2+a (\sum_{l'\in b(l)}p_{l'})^{-2}+\mu^2\Big{]}}
\prod_{l\in T^1}
e^{-\alpha_l\Big{[}(k+\sum_{l'\in b(l)}p_{l'})^2+a (k+\sum_{l'\in b(l)}p_{l'})^{-2}+\mu^2\Big{]}}.\nonumber\\
\eeqa

We Taylor-expand the last line. The odd terms in $p$ are zero after integration, as the branch momenta 
are linearly independent. For each term we have a development of the form
\beqa
e^{-\alpha_l\Big{[}(\sum_{l'\in b(l)}p_{l'})^2+a
(k+\sum_{l'\in b(l)}p_{l'})^{-2}+\mu^2\Big{]}} 
\bigl(1-\alpha_l k^2+\alpha_l^2k^4\int_{0}^{1}dt(1-t) e^{-t\alpha_l k^2}\bigr)
\eeqa
Using the multiscale bound \eqref{slicebound}, 
we see that collecting the first terms we get a bound like \eqref{powercounting}, thus a quadratic mass divergence. 
If we have at least one factor in $\alpha_l$ we gain at least $M^{-2i_l} \le M^{-2i}$ and we pay a factor
$k^2$ which is of order $M^{2e}$ because the external momenta is of scale $e$. Thus 
for all scales $j$ between $i$ and $e$ we have gained a factor $M^{-2}$ and the power
counting factor associated to the corresponding connected component 
$G^j_r$,  which was previously $M^{-(N(G^j_r) -4)}= M^{2}$, has become $M^{-(N(G^j_r)-2)} = 1$.
We get therefore a constant per slice as power counting for that connected component.
As usually we recognize here the logarithmically divergent wave function renormalization associated to $S$.
All other terms give convergent contributions, because a factor at least $M^{-4}$ per slice between $e$ and $i$
is gained.

\subsubsection{$2$-point function with two broken faces }

The amplitude of a one-particle irreducible $2$-point graph with two broken faces is
\beqa
{\cal A}(G^j_r)&=&\delta (k_1+k_2)
 \int  \prod_{l=1}^L d^4 p_l \int^{M^{-2(i_l-1)}}_{M^{-2i_l}}d\alpha_l~e^{-\alpha_l [p_l^2+ a p_l^{-2}+\mu^2]}\nonumber\\
&& \prod_{v\neq \bar v}\delta(q^v_1+q^v_2+q^v_3+q^v_3)
e^{\imath k_2\wedge(\sum_{l\in S} p_l)} \, ,
\eeqa
with $S\in {\cal L}$ the set of loop lines crossed by the second external line. Performing again the momentum routing, dropping the index in $k_2$ and the global $\delta$ function yields
\beqa
\label{2brokenfaces2point}
{\cal A}(G^j_r)&=&
 \int  \prod_{l=1}^{|{\cal L}|} d^4 p_l 
\int^{M^{-2(i_l-1)}}_{M^{-2i_l}}d\alpha_l~
\prod_{l\in {\cal L}}
e^{-\alpha_l [p_l^2+ a p_l^{-2}+\mu^2]}
\nonumber\\
&&
\prod_{l\in T^2}
e^{-\alpha_l\Big{[}(\sum_{l'\in b(l)}p_{l'})^2+a (\sum_{l'\in b(l)}p_{l'})^{-2}+\mu^2\Big{]}}
\nonumber\\
&&
\prod_{l\in T^1}
e^{-\alpha_l\Big{[}(k+\sum_{l'\in b(l)}p_{l'})^2+a
(k+\sum_{l'\in b(l)}p_{l'})^{-2}+\mu^2\Big{]}}~
e^{\imath k\wedge(\sum_{l\in S} p_l)} \,.
\eeqa

In order to establish the full dependence of the amplitude in the external momentum $k$ we divide all integrals over $p$ in two regions, $p<a^{1/4}$ and $p\ge a^{1/4}$. 
The integral over one $p$ in the region $p<a^{1/4}$ will count for $O(1)$ instead of $M^{4i}$ and using directly the power counting argument we bound  such a contribution to \eqref{2brokenfaces2point} by $M^{-2i}$ per slice, for all $k$.

We conclude that only the case with all $p$'s greater than $a^{1/4}$ can give rise to a non analytic behavior in $k$. In the following we will neglect all boundary terms on the sphere of radius $a^{1/4}$ as they are easy to bound uniformly in $k$.

We chose a line $l'\in S$, use
\beqa
\label{chestie}
e^{\imath k\wedge(\sum_{l\in S} p_l) }=-\frac{1}{k^2}\Delta_{p_{l'}} e^{\imath k\wedge(\sum_{l\in S} p_l) }
\eeqa
and integrate by parts in \eqref{2brokenfaces2point} to get
\beqa
&&{\cal A}(G^j_r)=-\frac{1}{k^2}
 \int  \prod_{l=1}^{|{\cal L}|} d^4 p_l e^{\imath k\wedge(\sum_{l\in S} p_l)} 
\int^{M^{-2(i_l-1)}}_{M^{-2i_l}}d\alpha_l~\Delta_{p_{l'}} \Big{(}
\prod_{l\in {\cal L}}
e^{-\alpha_l [p_l^2+ a p_l^{-2}+\mu^2]}\nonumber\\
&&\prod_{l\in T^2}
e^{-\alpha_l\Big{[}(\sum_{l'\in b(l)}p_{l'})^2+a (\sum_{l'\in b(l)}p_{l'})^{-2}+\mu^2\Big{]}}
\prod_{l\in T^1}
e^{-\alpha_l\Big{[}(k+\sum_{l'\in b(l)}p_{l'})^2+a
(k+\sum_{l'\in b(l)}p_{l'})^{-2}+\mu^2\Big{]}}
\Big{)}\,.
\eeqa
The derivatives acting on the Gaussian will give rise to insertions scaling like $\alpha$, $\alpha^2p^2$, 
$\alpha p^{-2}$, $\alpha p^{-4}$, $\alpha^{2}p^{-2}$, 
$ \alpha^{2}p^{-4}$. The first two terms scale as $M^{-2i}$ in a slice while the rest scale 
at least as $M^{-4i}$. Using again the trick \eqref{chestie},  and the power counting bound we get, 
when summing over all slices, a behavior like
\beqa
\label{2bound}
{\cal A}(G^j_r)=\frac{1}{k^4}\sum_{i=j}^{\infty}M^{-2i}=\frac{1}{k^2}\frac{M^{-2j}}{M^{2e}} K\, ,
\eeqa
with $K$ some constant, if $k\approx M^e$ (and consequently of scale $|e|$). As the scale $j$ is ultraviolet with respect to $|e|$ we bound
\beqa
\label{astuce}
M^{-2j-2e}\le M^{-2(|e|+e)}\le 1 \, .
\eeqa
We have thus proved that 
\beqa
\label{abdel}
{\cal A}(G^j_r)=\frac{1}{k^2}F(k)
\eeqa
with $F(k)$ a function uniformly bounded by a constant for all $k$.\footnote{In fact $F(k)$ is analytic in $k$, as it is a sum of absolutely convergent integrals of analytic functions in $\alpha$ and $k$.}
We identify the terms $F(0)$ as a {\it finite} renormalization for the coefficient $a$ in the Lagrangian. Note that using this scale decomposition 
there are {\it no} logarithmic subleading divergences for this  two point function with  two broken faces.

\subsection{Four-points  function}

The amplitude of a planar regular four-points graph is given by:
\beqa
{\cal A}(G^j_r)&=&\delta (k_1+k_2+k_3+k_4)e^{\frac{\imath}{2}\sum_{i<j}k_i\wedge k_j} \nonumber\\
&& \int  \prod_{l=1}^L d^4 p_l \int^{M^{-2(i_l-1)}}_{M^{-2i_l}}d\alpha_l~e^{-\alpha_l [p_l^2+ a p_l^{-2}+\mu^2]}
\prod_{v\neq \bar v}\delta(q^v_1+q^v_2+q^v_3+q^v_3) \, .
\eeqa
The first line reproduces exactly the Moyal four-points kernel.  Power counting leads to bound the second line 
by a constant per slice, thus it corresponds to a logarithmic divergence, which in turn generates a logarithmic coupling constant renormalization.

Some comments are in order for the planar four-points graphs with more than one broken face. 
Using \eqref{chestie} once, we get a bound like
\beqa
{\cal A}(G^j_r)=\frac{1}{k^2}\sum_{i=j}^{\infty}M^{-2i}=\frac{M^{-2j}}{M^{2e}} K\, ,
\eeqa
and by \eqref{astuce} we see that the amplitude of such a graph 
is a function of external momenta uniformly bounded by some constant.

\section{Conclusions and perspectives}
\renewcommand{\theequation}{\thesection.\arabic{equation}}   
\setcounter{equation}{0}

We have thus proved in this paper that the scalar model \eqref{revolutie}
is renormalizable at all orders in perturbation theory. The
renormalization of the planar regular graphs goes along the same lines as the
renormalization of the Euclidean $\phi^4$ on a $4-$dimensional commutative
space.  The non-planar graphs remain convergent
and the main difference concerns the planar irregular graphs. The
comparison with the action \eqref{act-normala}
(which is non-renormalizable, with UV/IR mixing) and with the
Grosse-Wulkenhaar model is summarized in the following table:

\begin{center}
\begin{tabular}{|l|c|c|c|c|c|c|}\hline

& \multicolumn{2}{|c|}{{\it model $\eqref{act-normala}$}}
& \multicolumn{2}{|c|}{{\it Grosse-Wulkenhaar model}}
& \multicolumn{2}{|c|}{{\it model $\eqref{revolutie}$}}\\ 
 & 2-points & 4-points & 2-points & 4-points & 2-points & 4-points\\
\hline
planar regular & ren. & ren. & ren. & ren. & ren. & ren.\\
\hline
planar irregular & UV/IR & ren. & conv. & conv. & finite ren. & convergent \\
\hline
non-planar & convergent & convergent & convergent & convergent & convergent & convergent \\
\hline
\end{tabular}
\end{center}

\bigskip
{\bf Acknowledgment:} We thank A. Abdesselam for indicating the proof of analicity of $F$ in \eqref{abdel}.  
Furthermore, Adrian Tanasa gratefully acknowledges the European Science Foundation Research Networking Program ``Quantum Geometry and Quantum Gravity'' for the Short Visit Grants 2219 and 2232.

\appendix
\section{Examples of graphs}
\renewcommand{\theequation}{\thesection.\arabic{equation}}   
\setcounter{equation}{0}
\label{exemple}

We illustrate the general results established in the previous section by some examples of two- and four-points graphs for which we analyze the Feynman amplitude.

\subsection{A two-point graph example}

Let us analyze the Feynman amplitude of the tadpole of Fig \ref{fig:tadpole}. This graph has $g=0$ but $B=2$. Due to the new renormalization group slices the parameter $\alpha$ of an internal line obeys $\alpha <\text{min}\{k^2,k^{-2} \} $. Therefore the amplitude of the non planar tadpole is (up to a constant)
\beqa
\label{t1}
\int_0^{\text{min}\{k^2,k^{-2} \}} d\alpha \int d^4p e^{ik\wedge p}e^{-\alpha (p^2+a p^{-2}+\mu^2)}.
\eeqa
applying \eqref{chestie} and integrating by parts holds
\beqa \label{naspa}
&&-\frac{1}{k^2} \int_0^{\text{min}\{k^2,k^{-2} \}} d\alpha \int d^4p e^{ik\wedge p}
\Delta_p e^{-\alpha (p^2+a p^{-2}+\mu^2)}
=\frac{1}{k^2}\int_0^{\text{min}\{k^2,k^{-2} \}} d\alpha \int d^4p e^{ik\wedge p} \nonumber\\
&&
\bigl(
8\alpha-\alpha^2(4p^2-8ap^{-2}+4a^2p^{-6})
\bigr)
e^{-\alpha (p^2+a p^{-2}+\mu^2)}\, .
\eeqa
All but the first and second terms in \eqref{naspa} can be bounded by $k^2$ when taking absolute values
such that there contribution to the amplitude of the tadpole is a constant. The coefficient of the $k^{-2}$ divergences is therefore
\beqa
\label{naspa2}
c=\int_0^{\text{min}\{k^2,k^{-2} \}} d\alpha \int d^4p e^{ik\wedge p}
\bigl(8\alpha+4 p^2 \alpha^2\bigr)
e^{-\alpha (p^2+a p^{-2}+\mu^2)}
\eeqa
Applying again \ref{chestie} and integrating again by parts holds only terms like
\beqa
c_n=\frac{1}{k^2}\int_0^{\text{min}\{k^2,k^{-2} \}} \alpha^2 d\alpha \int d^4p e^{ik\wedge p} 
(\alpha p^2)^n e^{-\alpha (p^2+a p^{-2}+\mu^2)}\, ,
\eeqa
with $n=0,1,2$. Taking absolute values, using $(\alpha p^2)^n e^{-\alpha p^2}<e^{-\frac{\alpha p^2}{2}}$ holds 
up to irrelevant constants
\beqa
c_n<\frac{1}{k^2}\int_0^{\text{min}\{k^2,k^{-2} \}} \alpha^2 d\alpha \frac{1}{\alpha^2}
=\frac{1}{k^2}\text{min}\{k^2,k^{-2} \}<1
\eeqa
We conclude that
\beqa
\label{rez}
\frac{1}{k^2}F(k) + G(k) \, ,
\eeqa
with $F$ and $G$ bounded and analytic at $k=0$.

\subsection{Planar irregular four-points graphs}

\begin{figure}[ht]
\centerline{\psfig{figure=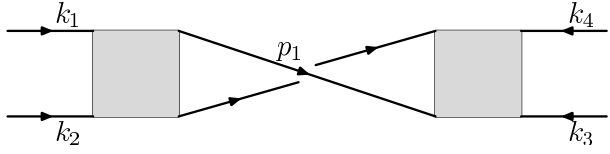,width=8cm}}
\caption{An example of a four-point Feynman graph, planar but with $2$ faces broken by external lines}
\label{fig:4}
\end{figure}

Take now the graph of Fig. \ref{fig:4}.  This graph has vanishing genus ($g=0$) and 
two faces broken by external lines ($B=2$).

The Feynman amplitude writes
\beqa
\label{alta2}
 \lambda^2 \int d^4 p_1 e^{-2i p_1\wedge (k_1+k_2)}
\frac{1}{p_1^2+
\mu^2+a\frac{1}{p_1^2}}
\, 
\frac{1}{(p_1+k_3+k_4)^2+
\mu^2+a\frac{1}{(p_1 +k_3 +k_4)^2}}
\eeqa
Let
\beqa
\label{notatii}
K&=&k_1+k_2 = - (k_3 + k_4),\nonumber\\
p_2&=&p_1+K.
\eeqa
We now deal with the integral \eqref{alta} as before, that is we use the Schwinger parametric representation and we express the oscillation factor using \eqref{chestie}. Integrating by parts as above, one has
\beqa
-\frac{\lambda^2}{K^2} \int_0^{\text{min}(K^2, K^{-2})}d\alpha_1 d\alpha_2
\int d^4 p_1 e^{-2i p_1\wedge K} \Delta_p 
\left[
e^{-\alpha_1 (p_1^2 +a p_1^{-2}+\mu^2)}e^{-\alpha_2 (p_2^2 +a p_2^{-2}+\mu^2)}
\right].
\eeqa
This further develops as:
\beqa
\label{naspa3}
&&-\frac{\lambda^2}{K^2} \int_0^{\text{min}(K^2, K^{-2})}d\alpha_1 d\alpha_2
\int d^4 p_1 e^{-2i p_1\wedge (k_1+k_2)} \nonumber\\
&&\left[
\left[-8 \alpha_1 + \alpha_1^2 (4 p_1^2 + \frac{4a^2}{p_1^6}-\frac{8a}{p_1^2})\right]
+\left[-8 \alpha_2 + \alpha_2^2 (4 p_2^2 + \frac{4a^2}{p_2^6}-\frac{8a}{p_2^2})\right]\right.\nonumber\\
&&\left. +
8\alpha_1 \alpha_2 
(p_{1\, \mu} - \frac{a}{p_1^4}p_{1\, \mu})(p_{2}^{\mu} - \frac{a}{p_2^4}p_{2}^{\mu})\right]
e^{-\alpha_1 (p_1^2 +a p_1^{-2}+\mu^2)}e^{-\alpha_2 (p_2^2 +a p_2^{-2}+\mu^2)}.
\eeqa
Note that some of the terms above are of the same type as the ones appearing in \eqref{naspa} and can be bounded by $K^2$ when taking absolute values. Thus, their contribution to the amplitude is a constant. The rest of  terms of \eqref{naspa3} can then be treated along the same lines as above. Take for example the second term of \eqref{naspa3}. This leads to an integral like
\beqa
\int_0^{\text{min}(K^2, K^{-2})}d\alpha_1 d\alpha_2 \alpha_1 \frac{1}{(\alpha_1 +\alpha_2)^2}
\eeqa
One performs first the definite integral on $\alpha_2$. This leads to two terms which can be easily bounded by $K^2$. 
Finally, one concludes that the integral \eqref{naspa3} leads to some constant result.

\end{document}